\documentclass{article}

\usepackage{arxiv}

\usepackage[utf8]{inputenc} 
\usepackage[T1]{fontenc}    
\usepackage{hyperref}       
\usepackage{url}            
\usepackage{booktabs}       
\usepackage{amsfonts}       
\usepackage{nicefrac}       
\usepackage{microtype}      
\usepackage{lipsum}
\usepackage{graphicx}
\graphicspath{ {./images/} }

\usepackage[utf8]{inputenc}
\usepackage[T1]{fontenc}
\usepackage{amsmath,amssymb,amsfonts}
\usepackage{multirow}
\usepackage{array}
\usepackage{xcolor}
\usepackage[skins,breakable]{tcolorbox}
\usepackage{xspace}
\usepackage{tikz}
\usetikzlibrary{positioning,arrows.meta,fit,backgrounds,calc,decorations.pathmorphing,shapes.geometric}
\definecolor{cObs}{HTML}{2C7FB8}
\definecolor{cPert}{HTML}{D7301F}
\definecolor{cTarg}{HTML}{6A51A3}
\definecolor{cOk}{HTML}{2CA25F}
\definecolor{cGray}{HTML}{9E9E9E}

\usepackage{enumitem}
\setlist[itemize]{nosep,itemsep=1pt,topsep=1pt,leftmargin=1.2em}
\setlist[enumerate]{nosep,itemsep=1pt,topsep=1pt,leftmargin=1.4em}

\newtcolorbox[auto counter]{finding}[1][]{
  colback=black!4!white, colframe=black!70!black, boxrule=0.7pt,
  left=4pt, right=4pt, top=3pt, bottom=3pt, breakable,
  before upper=\textbf{Finding \thetcbcounter.\ }, #1}

\newcommand{\So}{\mathcal{S}_{\mathrm{o}}}
\newcommand{\Sa}{\mathcal{S}_{\mathrm{a}}}
\newcommand{\St}{\mathcal{S}_{\mathrm{t}}}
\newcommand{\eps}{\varepsilon}

\newcommand{\model}{f_{\theta}}
\newcommand{\detname}{\emph{VetTraffic}\xspace}
\newcommand{\nbrop}{\mathrm{nbr}}
\newcommand{\loss}{\mathcal{L}}

\newcommand{\myparagraph}[1]{\smallskip\noindent\textbf{#1}:\xspace}

\usepackage[textsize=tiny,colorinlistoftodos,textwidth=20mm]{todonotes} 
\newif\ifhighlight
\highlighttrue
\newif\ifcomments
\commentstrue

\title{Revisiting the Adversarial Robustness of Graph-Based Traffic Forecasting}
\author{Qingzhao Zhang\\
The University of Arizoan\\
qzzhang@arizona.edu}

\begin{document}
\maketitle

\begin{abstract}
Traffic forecasting by graph-based AI is a critical component of intelligent transportation systems, motivating security research on robustness to malicious sensor readings. We argue that prior robustness evaluations are largely shaped by unrealistic threat models and untargeted objectives, so both attacks and defenses must be revisited.
We study a practical adversary with limited model knowledge and the ability to monitor and manipulate only a few road sensors. More importantly, practical attacks can be localized to specific links or routes, causing incorrect estimated arrival times or unnecessary rerouting while leaving the broader network largely unaffected. This targeted setting remains underexplored, and defenses such as adversarial training do not transfer well from the norm-bounded attacks they train on to structurally different, physics-aware attacks that mimic genuine congestion. We therefore reframe robustness as a detection problem, introducing a learned physics-informed detector whose output is fed to a hardened forecaster as an input feature and trained against adaptive attacks with the forecaster fixed.
We evaluate across a variety of model architectures and benchmarks. The physics-aware attack multiplies target-link error several-fold while the network-wide error barely moves, and adversarial training, tuned to norm-bounded perturbations, barely dents it. Our detection--mitigation defense improves even on adversarial training hardened against the physics-aware attack itself, on $13$ of $15$ model--dataset settings and by the widest margin on a held-out attack, at near-zero clean cost.
The results emphasize the need to examine abstracted AI adversarial attacks under application-specific constraints to assess their true security impacts.
\end{abstract}
\section{Introduction}

Traffic forecasting predicts near-future speeds or flows across a road network and
underpins navigation, signal control, and congestion management. State-of-the-art
forecasters are graph-based spatiotemporal models, most prominently spatiotemporal
graph neural networks (ST-GNNs), that jointly model
temporal dynamics and the road graph~\cite{stgcn,dcrnn,gwnet}. Because their errors
translate into routing and control decisions, their adversarial robustness
matters. Prior work shows that these forecasters can be fooled by perturbing sensor
inputs~\cite{advst,diffusion_attack}, and that \emph{adversarial training} (AT)
restores robustness, culminating in reinforced dynamic adversarial training
(RDAT)~\cite{rdat}. The prevailing message is that adversarially trained forecasters
are robust.

The reassurance rests on an unrealistic threat model. Prior evaluations grant the adversary a
large fraction of sensors under a norm budget and score a network-wide average that a localized
manipulation cannot move. A deployable adversary is more constrained yet more dangerous: limited
model knowledge, only the few sensors it can reach, partial observation, and a \emph{localized}
goal such as biasing one corridor to influence routing. Under these five axes, the resulting
attack shatters benchmark-robust forecasters at the targeted links while barely moving the global
metric, and adversarial training cannot close the gap, because a node's forecast is tied to its
own perturbable reading.

Worse, the attacker need not stay within a norm ball: it can shape a sparse perturbation to
resemble a genuine congestion event, spatially coherent and temporally smooth, an
on-manifold~\cite{stutz2019disentangling} attack a forecaster cannot reject
without also rejecting real congestion. Adversarial-training robustness is thus
\emph{attack-specific}, transferring only partially from the norm-bounded attack it trains on to
this congestion-shaped one.

We introduce \detname, a model-agnostic defense that \emph{vets} each sensor reading and
distrusts the corrupted ones rather than merely tolerating them, complementing adversarial
training. A physics-informed
detector scores each node by its inconsistency with graph neighbors and directed traffic
dynamics, and this suspicion signal is fed to the forecaster as an extra input feature, so the
forecaster keeps the trusted raw reading and learns to discount flagged nodes. Because the
detector learns to flag physics violations rather than to invert one specific attack, its signal
generalizes across attacks. We train the detector and a lightweight adapter against adaptive
attacks on top of a forecaster that is warm-started from an adversarially trained model and kept
\emph{frozen}.

We evaluate both the attacks and the defense across a range of forecasters and benchmarks.
A detection defense can look robust merely because the attacker fails to find a perturbation
rather than because none exists~\cite{carlini2017,athalye2018}, so we test against the
strongest adaptive adversary: fully white-box, detector-aware, and able to compute true
gradients through the defense. Against this attacker \detname\ improves on the strongest
adversarial-training baseline (hardening on the physics-aware attack \emph{itself}) on $13$ of
$15$ model--dataset cells, at no clean-accuracy cost, and by the widest margin on a held-out
attack it never trained against, where attack-specific hardening degrades but detection still
fires.

\myparagraph{Contributions}
\begin{itemize}
\item A five-axis threat model (limited knowledge, partial perturbation and observation, localized
targets, a realism bound) with a localized objective and matching attacks.
\item A physics-aware attack shaped like genuine congestion, and evidence that adversarial-training
robustness is \emph{attack-specific}.
\item \detname, a model-agnostic detection defense complementing adversarial training.
\item An adaptive evaluation showing detection generalizes across attacks, beating adversarial
training even when it is hardened on the attack itself, at near-zero clean cost.
\end{itemize}

\section{Background and Related Work}
\label{sec:background}

\subsection{Learning-based Traffic Forecasting}
A road network is a graph $G=(V,E)$ with $N$ sensor nodes; a forecaster $\model$ maps
a history window $X\in\mathbb{R}^{N\times F\times T_{\mathrm{in}}}$ (speed and
time-of-day) to the next $T_{\mathrm{out}}$ speeds $\hat Y=\model(X;A)$, scored by mean absolute error (MAE).
The state of the art is dominated by \emph{spatiotemporal GNNs} (ST-GNNs), e.g.,
STGCN~\cite{stgcn}, DCRNN~\cite{dcrnn}, and Graph WaveNet~\cite{gwnet}, with later work
adding spatial-temporal attention~\cite{astgcn,gman,sttn,staeformer}, learned or adaptive
adjacency~\cite{agcrn,mtgnn,staggcn}, and delay-aware or prompt-based universal
architectures~\cite{pdformer,unist}; we attack
\emph{five} such forecasters spanning the convolutional, spectral, diffusion-recurrent,
and attention families, and build our defense on Graph WaveNet and STGCN. The defense is not
ST-GNN-specific: it assumes only that a node is predictable from its graph neighbors, so it
wraps any forecaster.

\subsection{Adversarial Attacks on Traffic Forecasting}
Graph-based forecasters are vulnerable to input perturbations, crafted by spatiotemporal
saliency~\cite{advst}, diffusion~\cite{diffusion_attack}, or spatially focused
subsets~\cite{spatialattack}; \cite{wang2025transferability} further show black-box
\emph{transfer} poisoning across architectures. All evaluate a \emph{global}
degradation under an $\ell_\infty$ budget (Table~\ref{tab:related}); we reuse their
machinery but make the objective and metric \emph{localized} and replace the
$\ell_\infty$ ball with a physics-aware structural constraint (Section~\ref{sec:realistic}).

\subsection{Adversarial Training and Defenses}
Robustness for forecasters is pursued almost exclusively through adversarial
training~\cite{madry}: TRADES~\cite{trades} and RDAT~\cite{rdat}, the traffic-specific state of
the art, all harden against the global $\ell_\infty$ threat and add no detection. Detection and
purification appear only in other domains (Table~\ref{tab:related}): static graph
structure~\cite{gnnguard,prognn} and vision inputs~\cite{magnet,featuresqueeze}, often broken by
adaptive attacks~\cite{carlini2017,athalye2018}; neighbor imputation corrects faulty sensors as
data quality~\cite{coifman,boquet}. \detname\ is the only forecasting defense that turns
physics-informed detection into a forecaster feature and is verified against an adaptive attacker.

\begin{table}[t]
\centering\small
\caption{Difference from prior work by threat model (attacks above, defenses below).
WB/BB: white-/black-box; Adapt.: evaluated against an adaptive attacker; Anom.: distinguishes
attacks from genuine anomalies.}
\label{tab:related}
\begin{tabular}{lcccc}
\toprule
Attack & Knowl. & Perturbed & Objective & Bound \\
\midrule
AdvST$^{a}$     & BB & subset & global & $\ell_\infty$ \\
Diffusion$^{b}$ & WB & subset & global & $\ell_\infty$ \\
Spatial$^{c}$   & WB & subset & global & $\ell_\infty$ \\
\textbf{Ours}   & BB & few    & \textbf{local.} & \textbf{real.} \\
\bottomrule
\end{tabular}
\smallskip
\begin{tabular}{lcccc}
\toprule
Defense & Mechanism & Threat & Adapt. & Anom. \\
\midrule
PGD-AT$^{d}$     & AT           & global & --- & --- \\
TRADES$^{e}$     & AT           & global & --- & --- \\
RDAT$^{f}$       & AT           & global & --- & --- \\
GNNGuard$^{g}$   & detect       & graph  & --- & --- \\
Imputation$^{h}$ & impute       & faults & --- & \checkmark \\
\textbf{Ours}    & detect+feat. & \textbf{local.} & \checkmark & \checkmark \\
\bottomrule
\end{tabular}

{\footnotesize\raggedright \textbf{Note.} AT: adversarial training; defense threats
are $\ell_\infty$-bounded. $^{a}$\cite{advst}; $^{b}$\cite{diffusion_attack};
$^{c}$\cite{spatialattack}; $^{d}$\cite{madry}; $^{e}$\cite{trades};
$^{f}$\cite{rdat}; $^{g}$\cite{gnnguard}; $^{h}$\cite{coifman,boquet}.}
\end{table}

\section{Threat Model}
\label{sec:threat}

Prior robustness studies perturb an \emph{unconstrained} node set (often a large fraction of $V$)
under an $\ell_\infty$ budget $\|\delta\|_\infty\!\le\!\eps$ to maximize the \emph{global} error
$\mathrm{MAE}(\hat Y,Y)$. This is convenient but unrealistic: it over-grants the attacker's reach
and knowledge, its global objective matches no operational intent, and its $\ell_\infty$ bound is
inherited from image classification. We refine it into a deployable threat along five axes.

\myparagraph{Node sets ($\So,\Sa,\St$)} The attacker \textbf{observes} $\So$ (A3),
\textbf{perturbs} only $\Sa$ (A2), and is \textbf{scored} on the impact set $\St$ (A4), with
$\Sa\subseteq\So$ (control implies read access). The impact set may be a \emph{self-target}
($\St\subseteq\Sa$, moving a node's own forecast) or a \emph{neighbor-target}
($\St\cap\Sa=\varnothing$, reached through the graph). The remaining axes A1 and A5 limit knowledge
of $\model$ and the injectable value.

\myparagraph{(A1) Limited knowledge} Deployed forecasters are proprietary, so the attacker
does not know $\model$: it trains a surrogate on public data and transfers perturbations, which
succeed black-box across architectures~\cite{wang2025transferability}
(Figure~\ref{fig:transfer}), or issues bounded queries. White-box is thus only an upper bound.

\myparagraph{(A2) Partial perturbation} Corrupting a sensor requires physical or network access to
it, so rewriting $20\%$ of a city is implausible: a realistic attacker controls a small set
$\Sa\subseteq V$ and perturbs only there ($\delta_i=0\ \forall i\notin\Sa$). Both $|\Sa|$ and, by
(A5), the per-sensor values are constrained, shifting the difficulty onto \emph{which} links to
corrupt.

\myparagraph{(A3) Partial observation} With no global feed, the attacker sees a subset
$\So\subseteq V$ and infers the rest from the graph for node selection and surrogate calibration;
errors in the inferred state weaken naive attacks.

\myparagraph{(A4) Localized targets} A real adversary pursues a \emph{local} goal:
inflate the predicted travel time on a rival corridor, clear a route, or trigger a
reroute past a chosen location, not raise a network average. Objective and evaluation
are restricted to a target set $\St$:
\begin{equation}
\max_{\delta}\ \sum_{i\in\St} \ell\!\big(\model(X+\delta)_i,\,Y_i\big),
\quad
\mathrm{MAE}_{\St}=\tfrac{1}{|\St|}\!\sum_{i\in\St}\!|\hat Y_i-Y_i| .
\label{eq:local}
\end{equation}
We study $\St\subseteq\Sa$ (mislead a node's own forecast) and $\St\not\subseteq\Sa$
(mislead it through its neighbors). With $|\St|\ll N$, a large rise in
$\mathrm{MAE}_{\St}$ barely moves the global $\mathrm{MAE}$, so a global metric hides a
successful attack; robustness must be reported at $\St$.

\myparagraph{(A5) A capability-based bound} The $\ell_\infty$ ball, inherited from imperceptible
image perturbations, is the wrong constraint for sensors: the attacker does not add a small $\eps$,
it \emph{sets} the reported value within what the hardware permits. For the dominant inductive-loop
detector, contactless electromagnetic injection can fabricate or mask a vehicle detection at low
power~\cite{zhang2020emi,tu2019trick,kune2013ghost}, and spoofed feeds have already driven signal
control into congestion~\cite{chen2018congestion,weckert2020maps}. Two limits follow:
\emph{(i) range}: the reading stays within the sensor's dynamic range, a \emph{plausible} not
arbitrary value; \emph{(ii) coherence}: an isolated, impossible reading is trivially rejected, so
a stealthy attack must stay spatially and temporally coherent. We formalize this realism bound in
Section~\ref{sec:realistic}.

\section{Design}
\label{sec:design}
We instantiate the threat model of Section~\ref{sec:threat} as a concrete attack
(Section~\ref{sec:attack}) and the detection--mitigation defense it motivates
(Section~\ref{sec:defense}).
\subsection{Attack Design}
\label{sec:attack}

\begin{figure}[t]
\centering
\includegraphics[width=0.6\linewidth]{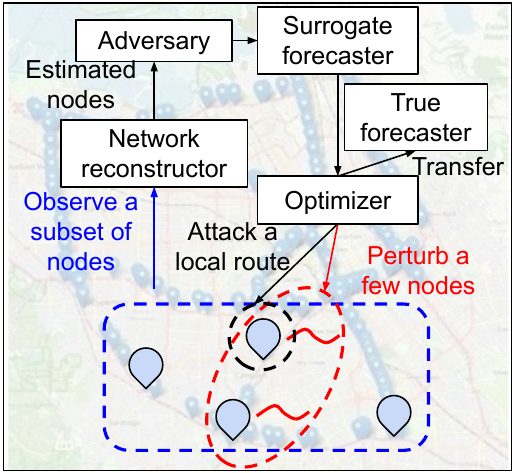}
\caption{Attack overview: observe $\So$, reconstruct the unobserved state, train a transferable
surrogate, then perturb the controlled sensors $\Sa$ to bias the target route $\St$.}
\label{fig:attack}
\end{figure}

Realizing the threat model of Section~\ref{sec:threat} takes four steps
(Figure~\ref{fig:attack}): \textbf{(1)}~select
the node sets ($\St,\So,\Sa$); \textbf{(2)}~reconstruct the unobserved state from $\So$;
\textbf{(3)}~train a surrogate to supply gradients without the deployed model; and
\textbf{(4)}~optimize the injected values on $\St$ by masked projected gradient descent (PGD).
Steps~1--3 handle the
attacker's limited knowledge and observation and are shared; step~4 carries the objective: the
physics-aware objective (Section~\ref{sec:realistic}), or, as an ablation, a norm-bounded baseline
(masked PGD on the localized objective~\eqref{eq:local} under $\|\delta\|_\infty\!\le\!\eps$).
Steps~1--3 use standard components, described next; the physics-aware objective is our novel part.

\myparagraph{Target and Access Selection}
\label{sec:partial}
The adversary's intent fixes the impact set $\St$; it then commits its physical budget to
the sensors it can observe and perturb, $\Sa\subseteq\So$. We define three node-selection
algorithms spanning realistic attacker capabilities:
\begin{itemize}[leftmargin=1.5em,itemsep=1pt,topsep=2pt]
\item \textbf{Degree.} Rank nodes by their graph degree and take the top $K$; uses only
the public road graph and needs no model access.
\item \textbf{Surrogate saliency.} Rank nodes by a calibrated sensitivity score
$\bar g_i=\mathbb{E}_{(X,Y)}\,|\partial\ell/\partial X_i|$ accumulated over a calibration
set on the surrogate, and take the top $K$; the strongest few-node choice.
\item \textbf{Corridor.} Grow a connected path of $K$ nodes by breadth-first search along
the strongest edges from a seed node; models a route-level target.
\end{itemize}
The same ranking selects which of the controlled sensors carry the perturbation.

\myparagraph{State Reconstruction under Partial Observation}
Because the attacker sees only the observed sensors $\So$, it must estimate the rest of
the network before scoring saliency or seeding the perturbation. We use two reconstruction
options, one closed-form and one learned.

\emph{Graph-Laplacian interpolation.} The unobserved values are filled by the
smoothest graph signal that agrees with the observations,
\begin{equation}
\hat X=\arg\min_{Z:\,Z_{\So}=X_{\So}}\ \operatorname{tr}\!\big(Z^{\!\top} L\,Z\big),
\qquad L=D-A,
\label{eq:recon}
\end{equation}
a closed-form harmonic extension that requires no training.

\emph{Learned reconstructor.} Alternatively, a GNN imputes masked nodes self-supervised
(hide a random subset, regress it from the rest and the graph); any imputation architecture
applies. Under sparse observation, reconstruction quality, not model access, limits the attack: a
learned reconstructor ties neighbour interpolation on dense-speed graphs and helps most where
observation is sparsest.

\myparagraph{Surrogate Training and Transfer}
Lacking the deployed forecaster, the attacker trains a surrogate $\hat\model$ (possibly a different
architecture) on public data and optimizes through it; a single surrogate already transfers, and an
ensemble sharpens the direction. The white-box attack we report unless noted is the transfer
attack's upper bound, keeping defense claims conservative; Section~\ref{sec:attack-exp} quantifies
the gap.

\myparagraph{Physics-Aware Attack}
\label{sec:realistic}
Our default attack drops the norm ball: a compromised sensor can report any physically valid
value, so what makes an injection evasive is not small magnitude but \emph{mimicking a genuine
anomaly}: a spatially coherent, temporally smooth slowdown rather than high-frequency noise. We
therefore drive the forecast toward a fabricated congested state while keeping the injection
consistent with the road graph and traffic dynamics, solving
\begin{equation}
\begin{aligned}
\min_{\delta:\,\mathrm{supp}(\delta)\subseteq\Sa}\
\sum_{i\in\St}&\ell\!\big(\model(X{+}\delta)_i,\,y^{\mathrm{jam}}\big)\\
&+\lambda_1\|\delta\|_1
+\lambda_2 R_{\mathrm{sp}}(X{+}\delta)\\
&+\lambda_3 R_{\mathrm{tm}}(X{+}\delta),
\end{aligned}
\label{eq:realistic}
\end{equation}
where $y^{\mathrm{jam}}$ is a fabricated congested (low-speed) target so that minimizing $\ell$
pushes the predicted speed down, $R_{\mathrm{sp}}(Z)=\sum_i|Z_i-\nbrop(Z)_i|$, with $\nbrop(Z)=PZ$ the neighbor average
($P$ the row-normalized adjacency), penalizes
inconsistency with graph neighbors (rewarding a coherently propagating slowdown), and
$R_{\mathrm{tm}}(Z)=\sum_{i,t}|Z_{i,t}-2Z_{i,t-1}+Z_{i,t-2}|$ penalizes temporal jerk (bounded
acceleration). The $\ell_1$ term keeps the injection sparse, on a few key links with a thin
consistency halo, while the magnitude is otherwise unconstrained. Unless noted, the default
setting perturbs the target corridor and its one-hop halo with weights $\lambda=(\ell_1\,0.02,\,
\mathrm{sp}\,1.0,\,\mathrm{tm}\,0.3)$ over $40$ steps; we vary these as impacting factors
(Section~\ref{sec:attack-exp}). The naming is deliberate: the attacker is physics-\emph{aware},
shaping the perturbation to hide, while the defender's detector (Section~\ref{sec:defense}) is
physics-\emph{informed}, using the same structure to reveal it. Being a valid input state rather
than a bounded perturbation, this is the attack the defenses of Section~\ref{sec:defense} must
confront.


\subsection{Defense Design}
\label{sec:defense}

\myparagraph{Key Insight}
\label{sec:def-insight}
Adversarial training is \emph{attack-specific}: even hardened on the physics-aware attack, it pays
an accuracy--robustness cost, since a congestion-shaped injection is a valid input state, so
discounting it also discounts genuine congestion. We add an orthogonal primitive: a \emph{detector}
that flags nodes inconsistent with traffic physics (a spoofed node deviates from its graph
neighbors and shows up in directed up/down-stream and temporal residuals), trained with its own
objective so it is \emph{attack-agnostic}. Because it learns \emph{what a physics violation looks
like} rather than how to invert one attack, its signal generalizes exactly where hardening
degrades. The question is how to \emph{use} it: overwriting suspicious inputs with a neighbor
estimate (sanitization) discards the raw reading and fails when the attack also perturbs neighbors,
so we instead feed the detector's evidence to the forecaster as \emph{extra features}, letting it
discount flagged nodes without a hard substitution.

\begin{figure}[t]
\centering
\includegraphics[width=0.95\linewidth]{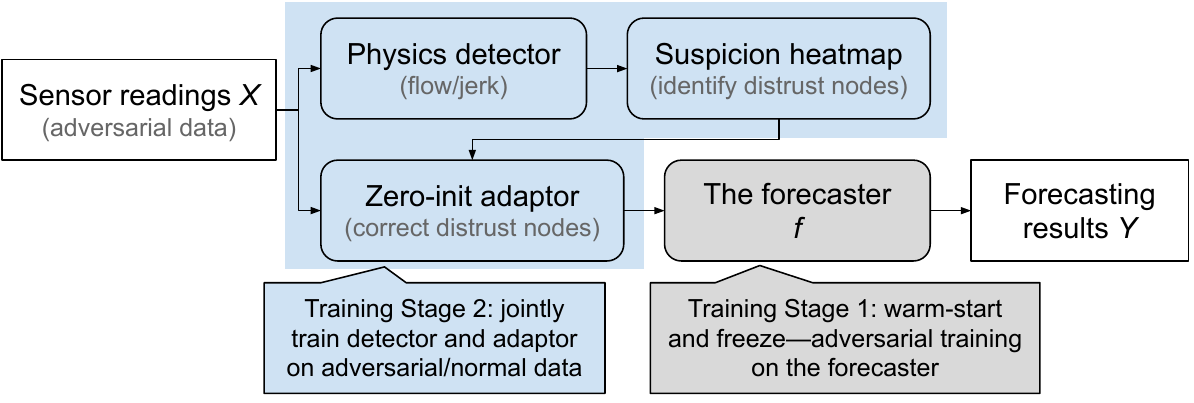}
\caption{\detname training stages. A physics-informed detector flags each
sensor by its inconsistency with directed flow and temporal dynamics (suspicion heatmap $s$); a
zero-init adapter feeds $s$ into the forecaster $f$ as an extra input feature. Stage~1 warm-starts
$f$ from AT-physics and freezes it (gray); Stage~2 trains the detector and adapter (blue).}
\label{fig:ncguard}
\end{figure}

\myparagraph{The \detname\ Design}
\label{sec:def-design}
\detname\ is a model-agnostic wrapper that turns a physics-informed detector into an input
feature for the forecaster (Figure~\ref{fig:ncguard}). It is designed as an \emph{increment} over
the strongest adversarial-training baseline: warm-started from an AT model and initialized, by a
zero-init adapter, to be \emph{identical} to it, so training only moves it where the detection
signal helps. A keep-best selection over training (below) keeps it from regressing below that
baseline on validation, though a fully adaptive attacker can still exploit the added component in
a minority of settings (Section~\ref{sec:def-exp}).

\emph{Neighbor and directed operators:} Let $P=D^{-1}A$ be the row-normalized adjacency
(a neighbor average) and $\nbrop(X)=PX$ the per-node estimate from neighbors. Directed variants
$P_{\mathrm{out}}{=}\mathrm{rownorm}(A)$ and $P_{\mathrm{in}}{=}\mathrm{rownorm}(A^{\!\top})$
separate downstream from upstream propagation. These drive both detection and the fused feature.

\detname\ comprises three components.

\myparagraph{(1) Physics-informed detector} For each node the detector consumes physics
residuals: its deviation from the directed up/down-stream estimates,
$r_{\mathrm{out}}{=}X{-}P_{\mathrm{out}}X$ and $r_{\mathrm{in}}{=}X{-}P_{\mathrm{in}}X$, and a
temporal-jerk term (bounded acceleration). Over a directed message-passing network it emits a
suspicion logit $\ell$ with $s=\sigma(\ell)\in[0,1]^N$. These residuals spike exactly where a
node disagrees with traffic dynamics, the fingerprint of a spoofed sensor, and are the same
structure the physics-aware attacker (Section~\ref{sec:realistic}) works to suppress: the
attacker is physics-\emph{aware} to hide, the detector physics-\emph{informed} to reveal.

\myparagraph{(2) Detection-as-feature via a zero-init adapter} Rather than overwrite the input,
we fuse the detector's evidence (the heatmap $s$, the neighbor estimate $PX$, and the directed
residuals) as extra node features:
\begin{equation}
X_{\mathrm{aug}} = X + \mathrm{adapter}\big([\,X,\ s,\ PX,\ r_{\mathrm{out}},\ r_{\mathrm{in}}\,]\big).
\label{eq:fuse}
\end{equation}
The adapter's output layer is \emph{zero-initialized}, so $X_{\mathrm{aug}}{=}X$ at the start and
the wrapper is exactly the underlying forecaster; training only moves it where the evidence helps.
The forecaster keeps the raw reading and leans on the neighbor evidence for flagged nodes, avoiding
the information loss of a hard substitution. Being differentiable, the adapter passes an adaptive
attacker true gradients, avoiding the \emph{gradient masking}~\cite{athalye2018} that makes
non-differentiable defenses only appear robust.

\myparagraph{(3) Forecaster: frozen AT baseline} The forecaster is warm-started from AT on the
physics-aware attack (the strongest baseline) and \emph{frozen}, so \detname\ only adds the
detection feature; its attack-agnostic signal helps most where the attack is one the AT model never
saw. Unfreezing $f$ to fine-tune adds a small further gain but is not required.

\myparagraph{Training} The detector and adapter are trained together while the forecaster stays
fixed. Each step draws $K$ target nodes, crafts an \emph{adaptive} physics-aware attack
$X_{\mathrm{adv}}$ through the full pipeline, and labels the perturbed nodes $m\in\{0,1\}^N$. We
keep the checkpoint with the lowest attacked validation error, starting from the zero-adapter
(AT-physics) state, so the adapter is retained only when it helps against the adaptive attack.
The loss combines forecasting on clean and attacked inputs with detection supervision:
\begin{align}
\loss =\ & \mathrm{MSE}(\model(X_{\mathrm{aug}}),Y)
      + \mathrm{MSE}(\model(X_{\mathrm{adv,aug}}),Y)\nonumber\\
    & + \lambda\big[\,\mathrm{BCE}(\ell_{\mathrm{adv}},m)
      + \mathrm{BCE}(\ell_{\mathrm{clean}},\mathbf{0})\,\big].
\label{eq:loss}
\end{align}
Gradients update only the detector and adapter; $\model$ stays frozen, so this is not
end-to-end training of the forecaster. The detection terms keep the signal attack-agnostic (firing
on physics violations, silent on clean nodes), so it transfers to unseen attacks, while the
zero-init adapter guarantees clean accuracy is preserved.

\section{Evaluation}
\label{sec:eval}
We evaluate the attack and the defense in turn on the same localized threat, reporting
target-corridor error alongside the network-wide metric.
\subsection{Attack Evaluation}
\label{sec:attack-exp}

\myparagraph{Experiment Setup}
We evaluate five graph-based spatiotemporal forecasters on three benchmarks.
\begin{itemize}[leftmargin=1.4em,itemsep=1pt,topsep=2pt]
\item \textbf{Models:} Graph WaveNet~\cite{gwnet} and STGCN~\cite{stgcn} (spatial and
spectral convolution), DCRNN~\cite{dcrnn} (recurrent diffusion), and STTN~\cite{sttn} and
STAGGCN~\cite{staggcn} (spatiotemporal attention). Graph WaveNet and STGCN are also the
forecasters we defend in Section~\ref{sec:def-exp}.
\item \textbf{Benchmarks:} PEMS-BAY ($325$ sensors) and METR-LA ($207$) report speed in mph~\cite{dcrnn};
PeMS-D4 ($307$) reports flow in veh/5\,min~\cite{staggcn}. History and horizon are $12$ steps.
\end{itemize}
We report the \emph{target-MAE} ($\mathrm{MAE}_{\St}$, also called target-corridor MAE below),
the mean absolute error over the attacked corridor in the dataset's native unit, and use the
network-wide MAE as a \emph{concealment} measure. All attacks
target a connected $K{=}5$-node corridor. Our \textbf{default attack is the physics-aware attack}
(Section~\ref{sec:realistic}, default weights, $40$ steps); we first characterize the norm-bounded
attack ($\eps{=}0.5$, $30$-step PGD) as the ablation baseline, then contrast it with the
physics-aware attack under defenses (Table~\ref{tab:realistic}).

\myparagraph{Main Results}
\label{sec:undef-eff}
Under the norm-bounded ablation, every architecture is severely degraded on the targeted corridor
while the network-wide error barely moves (Table~\ref{tab:undefended}): the manipulation is
confined to the corridor a routing or signal-control system reads, and is nearly invisible in
aggregate accuracy. Two effects are traffic-specific: sparse-\emph{flow} PeMS-D4, with fewer
redundant neighbours, amplifies every attack, and the recurrent DCRNN is most vulnerable. The
physics-aware attack (Table~\ref{tab:realistic}, \emph{Undef}) is stronger still, defeating
adversarial training as we show next.

\begin{table}[t]
\centering\small
\setlength{\tabcolsep}{5pt}
\caption{Norm-bounded attack (ablation): clean\,$\to$\,attacked target-MAE, native units.}
\label{tab:undefended}
\begin{tabular}{lccc}
\toprule
Model & METR-LA & PEMS-BAY & PeMS-D4\\
      & (mph) & (mph) & (veh/5\,min)\\
\midrule
Graph WaveNet & $4.0\!\to\!8.8$  & $1.9\!\to\!6.1$  & $6.2\!\to\!28.4$\\
STGCN         & $5.9\!\to\!27.0$ & $2.1\!\to\!8.2$  & $7.1\!\to\!90.5$\\
DCRNN         & $3.8\!\to\!26.4$ & $2.0\!\to\!29.0$ & $6.2\!\to\!152.7$\\
STTN          & $4.7\!\to\!10.3$ & $2.1\!\to\!5.5$  & $6.5\!\to\!35.6$\\
STAGGCN       & $6.5\!\to\!14.4$ & $2.1\!\to\!6.7$  & $6.3\!\to\!85.1$\\
\bottomrule
\end{tabular}
\end{table}

\begin{figure}[t]
\centering
\includegraphics[width=\linewidth]{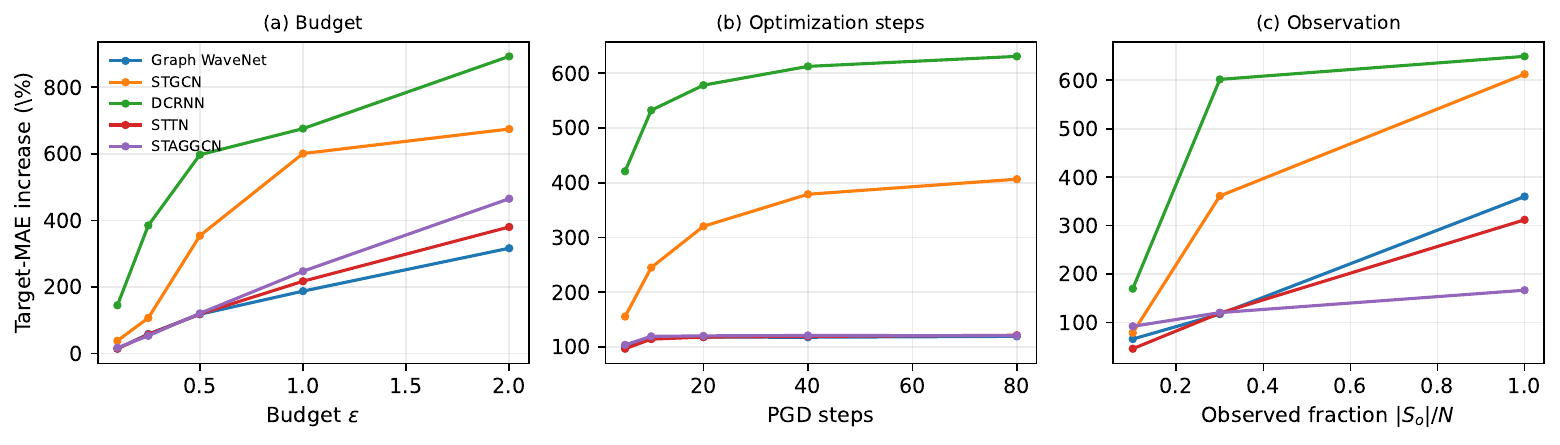}
\caption{Impacting factors of the localized attack (METR-LA), per model: target-MAE increase
versus (a)~budget $\eps$, (b)~PGD steps, and (c)~observed fraction $|\So|/N$.}
\label{fig:factors}
\end{figure}

\myparagraph{Impacting Factors of Attacks}
\label{sec:factors}
A one-factor sweep of $\sim\!400$ configurations (Figure~\ref{fig:factors}) shows attack strength
is governed by two attacker capabilities. \emph{Budget}: target-MAE rises with $\eps$ then
saturates once the corridor is maximally disrupted (Fig.~\ref{fig:factors}a). \emph{Steps}: it
rises with PGD steps and plateaus by $20$--$40$ (Fig.~\ref{fig:factors}b), the
monotone-then-saturating profile of a genuine optimizer, not one stalled by uninformative
gradients. \emph{Coverage}: effectiveness grows with the observed fraction
(Fig.~\ref{fig:factors}c), because reconstruction (Laplacian or neighbour interpolation, beating
zero-fill) improves the attacker's state estimate. Node choice matters least: ranking by public
graph \emph{degree} gives the largest error in three of four settings (Table~\ref{tab:nodesel})
and needs no model access, so surrogate saliency buys no consistent advantage.

\begin{table}[t]
\centering\small
\setlength{\tabcolsep}{5pt}
\caption{Node-selection strategy for the $K{=}5$ target--perturbation set: attacked
target-MAE (mph). Best per row in \textbf{bold}.}
\label{tab:nodesel}
\begin{tabular}{llcccc}
\toprule
Model & Data & Degree & Saliency & Corridor & Random\\
\midrule
GWNet & METR-LA  & \textbf{9.7} & 8.2 & 8.0 & 8.4\\
GWNet & PEMS-BAY & 5.2 & 6.3 & 6.2 & \textbf{7.0}\\
STGCN & METR-LA  & \textbf{40.2} & 38.3 & 38.3 & 29.7\\
STGCN & PEMS-BAY & \textbf{13.4} & 11.0 & 12.0 & 12.0\\
\bottomrule
\end{tabular}
\end{table}

\begin{finding}
The attacker's leverage is physical reach and observation, not compute or model access: a
model-free, degree-based target choice already rivals saliency-guided selection, so security
cannot rest on model secrecy or on making the attack harder to optimize.
\end{finding}

\begin{figure}[t]
\centering
\includegraphics[width=\linewidth]{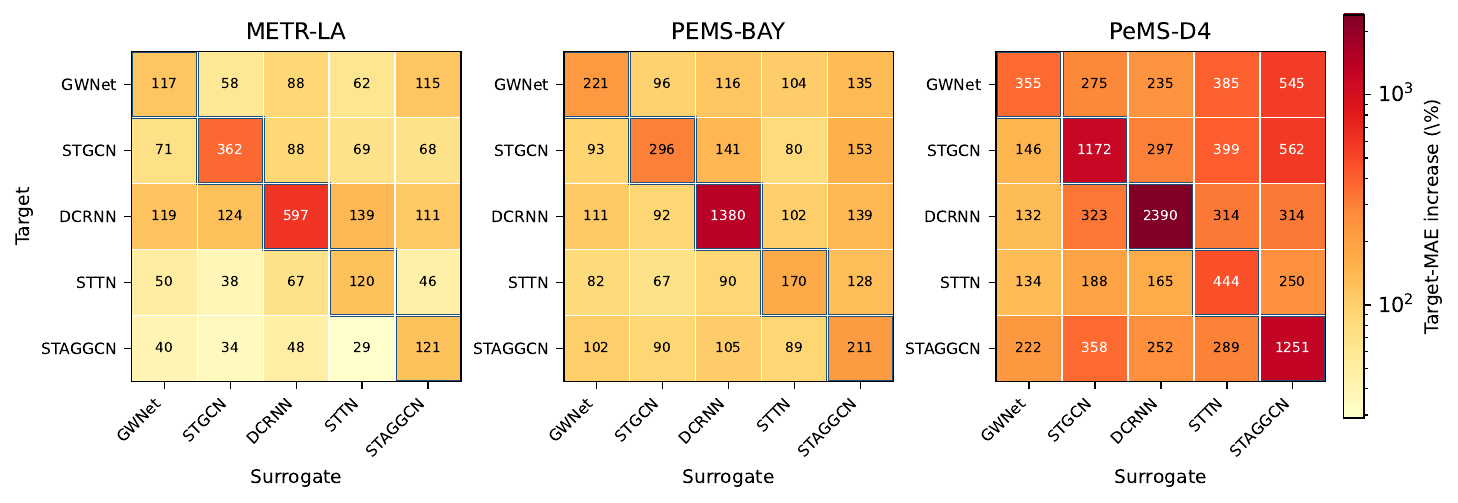}
\caption{Black-box transfer of the localized attack: target-MAE increase (\%, log color) for
every surrogate$\to$target pair per benchmark. Rows are the target, columns the surrogate; the
boxed diagonal is the white-box bound.}
\label{fig:transfer}
\end{figure}

\myparagraph{Black-Box Transfer}
\label{sec:transfer-exp}
Without the model, the adversary crafts on a surrogate of a different architecture and transfers,
reconstructing unseen nodes by Laplacian interpolation~\eqref{eq:recon}. The $5\times5\times3$
surrogate--target matrix (Figure~\ref{fig:transfer}) shows every off-diagonal surrogate still
inflates corridor error, below the white-box diagonal but nonzero, most on sparse-flow PeMS-D4 and
the recurrent DCRNN. Reconstruction matches full-observation transfer, so under partial
observation reconstruction quality, not model access, binds. We evaluate defenses white-box to
stay conservative.

\myparagraph{Against Adversarial Training}
\label{sec:attack-specific}
We now attack \emph{norm-trained} adversarial training (RDAT, the strongest such variant) under
the physics-aware attack (Table~\ref{tab:realistic}, \emph{Undef} vs.\ \emph{RDAT} columns) to
test whether hardening transfers across attack types.

Adversarial training is \emph{attack-specific}. It curbs the norm-bounded attack it trains on, but
against the physics-aware attack it is nearly useless: the attacked corridor error stays close to
undefended, and on sparse-flow PeMS-D4 it is often \emph{worse} ($145.0$ vs.\ $137.2$ undefended
on STAGGCN). This is the accuracy--robustness wall: a congestion-shaped injection is a valid input
state, so training to reject it also rejects genuine congestion. The defense must therefore
anticipate the attack: either by hardening on it directly (our AT-physics baseline) or, better,
by detecting it (Section~\ref{sec:def-exp}).

\begin{finding}
Adversarial training is attack-specific to a fault: hardened on one attack it can be \emph{worse}
than no defense on another. Robustness is earned per threat, not inherited across attack shapes.
\end{finding}

The defense must therefore either harden on the physics-aware attack directly or detect it;
Section~\ref{sec:def-exp} shows detection is the stronger of the two.

\subsection{Defense Evaluation}
\label{sec:def-exp}

\myparagraph{Experiment Setup}
One apples-to-apples harness: identical batches, the same target corridor, and absolute
target-corridor MAE (not percentages over differing clean baselines). We compare four defenses:
\begin{itemize}[leftmargin=1.4em,itemsep=1pt,topsep=2pt]
\item \textbf{Undef}: no defense.
\item \textbf{RDAT}: adversarial training on the \emph{norm-bounded} attack (the traffic-specific
state of the art).
\item \textbf{AT-physics}: adversarial training on the \emph{physics-aware} attack itself, the
strongest baseline, as it hardens on the very attack it faces.
\item \textbf{\detname} (Ours): the detection feature on top of a frozen AT-physics forecaster.
\end{itemize}
under two attack settings, both white-box and crafted adaptively through the full pipeline:
\begin{itemize}[leftmargin=1.4em,itemsep=1pt,topsep=2pt]
\item \textbf{Matched}: the physics-aware attack at the default the defense trains against.
\item \textbf{Held-out}: the same attack widened to a two-hop halo, a reach never trained
on, the key test of whether detection \emph{generalizes}.
\end{itemize}
A norm-bounded control ties across defenses (AT-physics absorbs it via RDAT) and is omitted.

\myparagraph{Main Results}
\detname\ improves on AT-physics on $13$ of $15$ cells under both attacks
(Table~\ref{tab:realistic}), cutting targeted error by up to $6.7$ (matched) and $10.6$
(held-out), at no clean-accuracy cost; on the three seeded cells the margins are stable
(std $\le 0.4$, each ${>}2\sigma$). The two exceptions are both on METR-LA, the noisiest graph:
STGCN/METR-LA ties (the keep-best guard falls back to AT-physics), and STTN/METR-LA regresses
slightly ($-0.9$) where the adaptive attacker exploits the adapter beyond what validation captures.

\begin{finding}
Distrust beats tolerance: handing a \emph{frozen} forecaster a signal of \emph{where} to distrust
its inputs outperforms adversarially retraining the forecaster on the attack itself, at no
clean-accuracy cost.
\end{finding}

\begin{table}[t]
\centering\small
\setlength{\tabcolsep}{3pt}
\caption{Attack vs.\ defenses on one harness (target-corridor MAE, native units).
AT${=}$AT-physics, Ours${=}$\detname; \textbf{bold} where Ours improves on AT ($13/15$).}
\label{tab:realistic}
\begin{tabular}{lcccccc}
\toprule
 & \multicolumn{4}{c}{Physics-aware} & \multicolumn{2}{c}{Held-out}\\
\cmidrule(lr){2-5}\cmidrule(lr){6-7}
Model/Data & Undef & RDAT & AT & Ours & AT & Ours\\
\midrule
GWNet/BAY & 41.1 & 39.7 & 14.3 & \textbf{11.9} & 16.0 & \textbf{13.2}\\
GWNet/LA & 30.5 & 28.5 & 29.3 & \textbf{26.6} & 29.5 & \textbf{27.2}\\
GWNet/D4 & 61.5 & 77.2 & 36.5 & \textbf{32.2} & 37.8 & \textbf{33.1}\\
STGCN/BAY & 47.3 & 45.5 & 13.7 & \textbf{12.6} & 25.6 & \textbf{22.3}\\
STGCN/LA & 29.7 & 33.0 & 26.4 & 26.4 & 26.5 & 26.5\\
STGCN/D4 & 64.8 & 65.9 & 26.6 & \textbf{19.9} & 32.3 & \textbf{21.7}\\
DCRNN/BAY & 56.5 & 48.7 & 17.2 & \textbf{15.0} & 31.4 & \textbf{26.3}\\
DCRNN/LA & 41.2 & 32.9 & 25.9 & \textbf{20.0} & 28.0 & \textbf{23.2}\\
DCRNN/D4 & 102.5 & 74.7 & 26.2 & \textbf{24.2} & 42.0 & \textbf{35.8}\\
STTN/BAY & 40.2 & 20.4 & 7.1 & \textbf{5.6} & 21.0 & \textbf{15.6}\\
STTN/LA & 29.3 & 30.7 & 20.9 & 21.8 & 23.6 & 23.8\\
STTN/D4 & 36.9 & 13.6 & 17.5 & \textbf{17.2} & 21.1 & \textbf{20.4}\\
STAGGCN/BAY & 76.4 & 72.0 & 34.2 & \textbf{27.1} & 40.4 & \textbf{33.2}\\
STAGGCN/LA & 105.0 & 97.2 & 59.3 & \textbf{50.0} & 61.5 & \textbf{52.3}\\
STAGGCN/D4 & 137.2 & 145.0 & 32.8 & \textbf{32.4} & 34.4 & \textbf{34.1}\\
\bottomrule
\end{tabular}
\end{table}

\myparagraph{Detection Generalizes Where Hardening Does Not}
The advantage is largest on the held-out attack. Tuned to one attack shape, AT-physics degrades
steeply as the attack widens beyond the one-hop halo it trained on (Table~\ref{tab:gencurve};
STGCN/PEMS-BAY $13.7\!\to\!34.8$ from one- to three-hop). The detector flags \emph{physics
violations} rather than inverting one attack, so it still fires on the widened attack and
\detname's margin \emph{grows} with the mismatch, up to $+6.2$ at three hops.

\begin{finding}
Detection generalizes where hardening cannot, and its edge \emph{grows} as the attack drifts from
training, the opposite of how hardened models degrade off-distribution, because it learns the
invariant signature of a spoof, not one attack's correction.
\end{finding}

\begin{table}[t]
\centering\small
\setlength{\tabcolsep}{4pt}
\caption{Generalization: attacked MAE (AT-physics\,/\,\detname) as the attack widens past the
trained one-hop halo; the margin grows.}
\label{tab:gencurve}
\begin{tabular}{lccc}
\toprule
Reach & GWNet/BAY & STGCN/BAY & GWNet/D4\\
\midrule
1-hop & 14.4\,/\,11.7 & 13.7\,/\,12.6 & 36.7\,/\,32.0\\
2-hop & 16.1\,/\,13.2 & 25.6\,/\,22.3 & 38.1\,/\,33.1\\
3-hop & 18.8\,/\,15.3 & 34.8\,/\,32.8 & 43.2\,/\,37.0\\
\bottomrule
\end{tabular}
\end{table}

\myparagraph{Ablation} Table~\ref{tab:defablation} isolates each piece. Every variant beats
AT-physics, but the \emph{physics-informed} detector earns the margin on dense-speed graphs
(GWNet/PEMS-BAY $13.8\!\to\!11.8$ vs.\ plain); on sparse-flow PeMS-D4 a plain detector suffices.
Freezing suffices; Stage-2 changes nothing, so \detname\ adds no forecaster training over AT-physics.

\begin{table}[t]
\centering\small
\setlength{\tabcolsep}{4pt}
\caption{Component ablation (matched attack, MAE). Physics detector helps most on dense graphs;
freezing suffices.}
\label{tab:defablation}
\begin{tabular}{lccc}
\toprule
Configuration & GWNet/BAY & STGCN/BAY & GWNet/D4\\
\midrule
AT-physics (no det.) & 14.2 & 13.7 & 36.6\\
\;Ours, plain det. & 13.8 & 13.3 & 31.2\\
\;Ours, scalar feat. & 12.4 & 12.3 & 31.2\\
\;Ours (phys., rich) & 11.8 & 12.6 & 32.1\\
\;+ Stage-2 & 11.9 & 12.5 & 32.7\\
\bottomrule
\end{tabular}
\end{table}

\myparagraph{Adaptive attacker} We stress the defense two ways~\cite{carlini2017,athalye2018}. A
\emph{detector-aware} attacker that also minimizes suspicion gains little: it lifts \detname's
error only from $11.9$ to $13.6$ (GWNet/PEMS-BAY), still below AT-physics ($14.2$), since an
injection strong enough to move the forecast cannot stay unflagged. And attacked error rises
\emph{monotonically} with steps ($5.1$ to $23.8$ over $10$--$80$), ruling out gradient masking
through the adapter.

\myparagraph{Genuine congestion} It also tells attacks from real slowdowns: on the lowest-speed
quartile of clean inputs the flag rate is $0\%$ (suspicion $0.01$) versus $62$--$100\%$ on attacks,
separating them at AUC $0.996$--$1.0$. A coherent slowdown is physically consistent; an injection
strong enough to matter is not.

\subsection{Case Study: Spoofing a Freeway Corridor}
\label{sec:casestudy}

\begin{figure}[t]
\centering
\includegraphics[width=\linewidth]{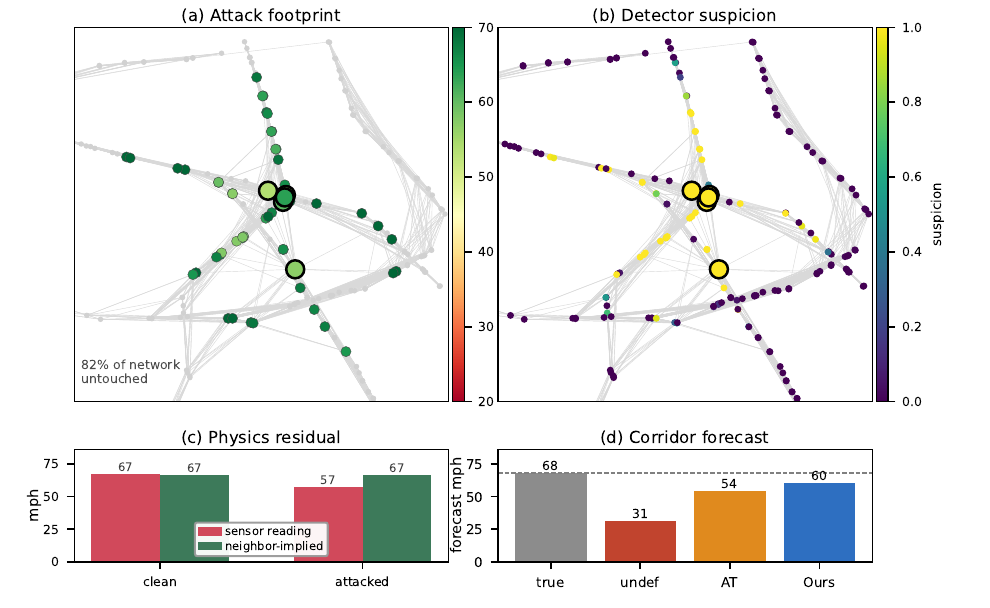}
\caption{Case study on PEMS-BAY (GWNet), real sensor geography (numbers in text).
\textbf{(a)}~attack footprint: the five-sensor corridor ($\St$, outlined) and its one-hop halo,
shaded by injected speed; \textbf{(b)}~detector suspicion, bright only on the corridor;
\textbf{(c)}~at a corridor node the injected reading falls below its directed-neighbor
estimate, the residual the detector flags; \textbf{(d)}~corridor forecast under each defense vs.\
the true speed.}
\label{fig:casestudy}
\end{figure}

One instance makes the mechanism concrete (Figure~\ref{fig:casestudy}) on PEMS-BAY. The adversary
injects a \emph{mild} coherent slowdown over a five-sensor corridor a routing service reads and its
one-hop halo: the spoofed readings drop only from $68.7$ to $58.6$~mph (true speed $68.1$), reading
as incipient congestion, not tampering. Yet undefended, the forecast collapses to a $31$~mph jam,
and adversarial training \emph{on this very attack} (AT-physics) still drops to $53.9$~mph, since a
valid-looking state anchored to the node's own reading cannot be rejected without also rejecting
real congestion. On a five-mile corridor that inflates travel time by $26\%$, enough to trigger
rerouting, while the network-wide error barely moves. \detname\ instead flags the corridor:
suspicion jumps $0\!\to\!1$ because the reading sits far below what its \emph{directed} neighbors
imply and is temporally over-smooth, so it holds the forecast at $60.3$~mph, halving the impact.
Detection need not \emph{tolerate} the spoofed reading, only \emph{distrust} it.

\section{Discussion}
\label{sec:discussion}

\myparagraph{Limitations} (i) \detname's gain is bounded by how discriminable a spoof is from
genuine dynamics: where the physics signal is weak or the graph is noisy (the METR-LA cells), the
fused feature helps little, so that regime stays with adversarial training. (ii) Although initialized
to match its AT baseline and kept there by a keep-best selection, a fully adaptive attacker through
the added adapter can still exploit it on a minority of settings (one of fifteen regresses
slightly); a construction that provably never regresses remains open. (iii) Detection supervision
assumes perturbed nodes can be labeled at training time, a genuine event mimicking the adversarial
signature could be misflagged, and graph-structure attacks remain out of scope.

\section{Conclusion}

We revisited graph-based traffic forecasting under a practical threat model and found its
reassuring robustness to be an artifact of an idealized attacker and a global metric: adversarially
trained forecasters degrade sharply at the nodes that matter, and their robustness is
\emph{attack-specific}. \detname\ generalizes better than hardening at negligible clean cost.

\bibliographystyle{unsrt} 
\bibliography{references}

@inproceedings{stutz2019disentangling,
  title={Disentangling adversarial robustness and generalization},
  author={Stutz, David and Hein, Matthias and Schiele, Bernt},
  booktitle={Proceedings of the IEEE/CVF conference on computer vision and pattern recognition},
  pages={6976--6987},
  year={2019}
}

@inproceedings{flow,
  title={Flow: A modular learning framework for mixed autonomy traffic},
  author={Wu, Cathy and Kreidieh, Abdul Rahman and Parvate, Kanaad and Vinitsky, Eugene and Bayen, Alexandre M},
  journal={IEEE Transactions on Robotics},
  volume={38},
  number={2},
  pages={1270--1286},
  year={2021},
  publisher={IEEE}
}

@inproceedings{stgcn,
  title={Spatio-temporal graph convolutional networks: a deep learning framework for traffic forecasting},
  author={Yu, Bing and Yin, Haoteng and Zhu, Zhanxing},
  booktitle={Proceedings of the 27th International Joint Conference on Artificial Intelligence},
  pages={3634--3640},
  year={2018}
}

@inproceedings{dcrnn,
  title={Diffusion Convolutional Recurrent Neural Network: Data-Driven Traffic Forecasting},
  author={Li, Yaguang and Yu, Rose and Shahabi, Cyrus and Liu, Yan},
  booktitle={International Conference on Learning Representations},
  year={2018}
}

@article{sttn,
  title={Spatial-temporal transformer networks for traffic flow forecasting},
  author={Xu, Mingxing and Dai, Wenrui and Liu, Chunmiao and Gao, Xing and Lin, Weiyao and Qi, Guo-Jun and Xiong, Hongkai},
  journal={arXiv preprint arXiv:2001.02908},
  year={2020}
}

@inproceedings{gwnet,
  title={Graph wavenet for deep spatial-temporal graph modeling},
  author={Wu, Zonghan and Pan, Shirui and Long, Guodong and Jiang, Jing and Zhang, Chengqi},
  booktitle={Proceedings of the 28th International Joint Conference on Artificial Intelligence},
  pages={1907--1913},
  year={2019}
}

@inproceedings{staggcn,
  title={Spatiotemporal adaptive gated graph convolution network for urban traffic flow forecasting},
  author={Lu, Bin and Gan, Xiaoying and Jin, Haiming and Fu, Luoyi and Zhang, Haisong},
  booktitle={Proceedings of the 29th ACM International conference on information \& knowledge management},
  pages={1025--1034},
  year={2020}
}

@inproceedings{astgcn,
  title={Attention based spatial-temporal graph convolutional networks for traffic flow forecasting},
  author={Guo, Shengnan and Lin, Youfang and Feng, Ning and Song, Chao and Wan, Huaiyu},
  booktitle={Proceedings of the AAAI conference on artificial intelligence},
  volume={33},
  number={01},
  pages={922--929},
  year={2019}
}

@inproceedings{gman,
  title={Gman: A graph multi-attention network for traffic prediction},
  author={Zheng, Chuanpan and Fan, Xiaoliang and Wang, Cheng and Qi, Jianzhong},
  booktitle={Proceedings of the AAAI conference on artificial intelligence},
  volume={34},
  number={01},
  pages={1234--1241},
  year={2020}
}

@inproceedings{agcrn,
  title={Adaptive graph convolutional recurrent network for traffic forecasting},
  author={Bai, Lei and Yao, Lina and Li, Can and Wang, Xianzhi and Wang, Can},
  journal={Advances in neural information processing systems},
  volume={33},
  pages={17804--17815},
  year={2020}
}

@inproceedings{mtgnn,
  title={Connecting the dots: Multivariate time series forecasting with graph neural networks},
  author={Wu, Zonghan and Pan, Shirui and Long, Guodong and Jiang, Jing and Chang, Xiaojun and Zhang, Chengqi},
  booktitle={Proceedings of the 26th ACM SIGKDD international conference on knowledge discovery \& data mining},
  pages={753--763},
  year={2020}
}

@inproceedings{pdformer,
  title={Pdformer: Propagation delay-aware dynamic long-range transformer for traffic flow prediction},
  author={Jiang, Jiawei and Han, Chengkai and Zhao, Wayne Xin and Wang, Jingyuan},
  booktitle={Proceedings of the AAAI conference on artificial intelligence},
  volume={37},
  number={4},
  pages={4365--4373},
  year={2023}
}

@inproceedings{unist,
  title={Unist: A prompt-empowered universal model for urban spatio-temporal prediction},
  author={Yuan, Yuan and Ding, Jingtao and Feng, Jie and Jin, Depeng and Li, Yong},
  booktitle={Proceedings of the 30th ACM SIGKDD conference on knowledge discovery and data mining},
  pages={4095--4106},
  year={2024}
}

@inproceedings{staeformer,
  title={Spatio-temporal adaptive embedding makes vanilla transformer sota for traffic forecasting},
  author={Liu, Hangchen and Dong, Zheng and Jiang, Renhe and Deng, Jiewen and Deng, Jinliang and Chen, Quanjun and Song, Xuan},
  booktitle={Proceedings of the 32nd ACM international conference on information and knowledge management},
  pages={4125--4129},
  year={2023}
}

@article{spatialattack,
  title={Spatially Focused Attack against Spatiotemporal Graph Neural Networks},
  author={Liu, Fuqiang and Miranda-Moreno, Luis F. and Sun, Lijun},
  journal={arXiv preprint arXiv:2109.04608},
  year={2021}
}

@inproceedings{rdat,
  title={Robust spatiotemporal traffic forecasting with reinforced dynamic adversarial training},
  author={Liu, Fan and Zhang, Weijia and Liu, Hao},
  booktitle={Proceedings of the 29th ACM SIGKDD Conference on Knowledge Discovery and Data Mining},
  pages={1417--1428},
  year={2023}
}

@article{diffusion_attack,
  title={Adversarial diffusion attacks on graph-based traffic prediction models},
  author={Zhu, Lyuyi and Feng, Kairui and Pu, Ziyuan and Ma, Wei},
  journal={IEEE Internet of Things Journal},
  volume={11},
  number={1},
  pages={1481--1495},
  year={2023},
  publisher={IEEE}
}

@inproceedings{advst,
 author = {Liu, Fan and Liu, Hao and Jiang, Wenzhao},
 booktitle = {Advances in Neural Information Processing Systems},
 editor = {S. Koyejo and S. Mohamed and A. Agarwal and D. Belgrave and K. Cho and A. Oh},
 pages = {19035--19047},
 publisher = {Curran Associates, Inc.},
 title = {Practical Adversarial Attacks on Spatiotemporal Traffic Forecasting Models},
 volume = {35},
 year = {2022}
}

@inproceedings{madry,
  title={Towards Deep Learning Models Resistant to Adversarial Attacks},
  author={Madry, Aleksander and Makelov, Aleksandar and Schmidt, Ludwig and Tsipras, Dimitris and Vladu, Adrian},
  booktitle={International Conference on Learning Representations (ICLR)},
  year={2018}
}

@inproceedings{trades,
  title={Theoretically Principled Trade-off between Robustness and Accuracy},
  author={Zhang, Hongyang and Yu, Yaodong and Jiao, Jiantao and Xing, Eric P and El Ghaoui, Laurent and Jordan, Michael I},
  booktitle={International Conference on Machine Learning (ICML)},
  year={2019}
}

@inproceedings{gnnguard,
  title={{GNNGuard}: Defending Graph Neural Networks against Adversarial Attacks},
  author={Zhang, Xiang and Zitnik, Marinka},
  booktitle={Advances in Neural Information Processing Systems (NeurIPS)},
  year={2020}
}

@inproceedings{prognn,
  title={Graph Structure Learning for Robust Graph Neural Networks},
  author={Jin, Wei and Ma, Yao and Liu, Xiaorui and Tang, Xianfeng and Wang, Suhang and Tang, Jiliang},
  booktitle={Proceedings of the 26th ACM SIGKDD Conference on Knowledge Discovery and Data Mining (KDD)},
  year={2020}
}

@inproceedings{magnet,
  title={{MagNet}: A Two-Pronged Defense against Adversarial Examples},
  author={Meng, Dongyu and Chen, Hao},
  booktitle={Proceedings of the ACM SIGSAC Conference on Computer and Communications Security (CCS)},
  year={2017}
}

@inproceedings{featuresqueeze,
  title={Feature Squeezing: Detecting Adversarial Examples in Deep Neural Networks},
  author={Xu, Weilin and Evans, David and Qi, Yanjun},
  booktitle={Network and Distributed System Security Symposium (NDSS)},
  year={2018}
}

@inproceedings{carlini2017,
  title={Adversarial Examples Are Not Easily Detected: Bypassing Ten Detection Methods},
  author={Carlini, Nicholas and Wagner, David},
  booktitle={Proceedings of the 10th ACM Workshop on Artificial Intelligence and Security (AISec)},
  year={2017}
}

@inproceedings{athalye2018,
  title={Obfuscated Gradients Give a False Sense of Security: Circumventing Defenses to Adversarial Examples},
  author={Athalye, Anish and Carlini, Nicholas and Wagner, David},
  booktitle={International Conference on Machine Learning (ICML)},
  year={2018}
}

@article{coifman,
  title={Detecting Errors and Imputing Missing Data for Single-Loop Surveillance Systems},
  author={Chen, Chao and Kwon, Jaimyoung and Rice, John and Skabardonis, Alexander and Varaiya, Pravin},
  journal={Transportation Research Record},
  volume={1855},
  number={1},
  pages={160--167},
  year={2003}
}

@article{boquet,
  title={A Variational Autoencoder Solution for Road Traffic Forecasting Systems: Missing Data Imputation, Dimension Reduction, Model Selection and Anomaly Detection},
  author={Boquet, Guillem and Morell, Antoni and Serrano, Javier and Vicario, Jos{\'e} L{\'o}pez},
  journal={Transportation Research Part C: Emerging Technologies},
  volume={115},
  pages={102622},
  year={2020}
}

@inproceedings{chen2018congestion,
  title={Exposing Congestion Attack on Emerging Connected Vehicle based Traffic Signal Control},
  author={Chen, Qi Alfred and Yin, Yucheng and Feng, Yiheng and Mao, Z. Morley and Liu, Henry X.},
  booktitle={Network and Distributed System Security Symposium (NDSS)},
  year={2018}
}

@misc{weckert2020maps,
  title={Google Maps Hacks},
  author={Weckert, Simon},
  year={2020},
  note={Art intervention: 99 smartphones fabricated a live traffic jam on Google Maps}
}

@inproceedings{kune2013ghost,
  title={Ghost Talk: Mitigating {EMI} Signal Injection Attacks against Analog Sensors},
  author={Kune, Denis Foo and Backes, John and Clark, Shane S and Kramer, Daniel and Reynolds, Matthew and Fu, Kevin and Kim, Yongdae and Xu, Wenyuan},
  booktitle={IEEE Symposium on Security and Privacy (S\&P)},
  year={2013}
}

@inproceedings{tu2019trick,
  title={Trick or Heat? Manipulating Critical Temperature-Based Control Systems Using Rectification Attacks},
  author={Tu, Yazhou and Rampazzi, Sara and Hao, Bin and Rodriguez, Angel and Fu, Kevin and Hei, Xiali},
  booktitle={ACM SIGSAC Conference on Computer and Communications Security (CCS)},
  year={2019}
}

@inproceedings{zhang2020emi,
  title={Detection of Electromagnetic Interference Attacks on Sensor Systems},
  author={Zhang, Youqian and Rasmussen, Kasper},
  booktitle={IEEE Symposium on Security and Privacy (S\&P)},
  year={2020}
}

@article{wang2025transferability,
  title={Transferability in Data Poisoning Attacks on Spatiotemporal Traffic Forecasting Models},
  author={Wang, Xin and Wang, Feilong and Hong, Yuan and Ban, Xuegang (Jeff)},
  journal={Transportation Research Part C: Emerging Technologies},
  year={2025}
}

\end{document}